\documentclass[preprint,aps,pre]{revtex4-1}

\usepackage[utf8]{inputenc}
\usepackage[T1]{fontenc}
\usepackage[english]{babel}
\usepackage{graphicx}
\usepackage{amssymb}
\usepackage{amsmath}
\usepackage{setspace}
\usepackage{array}
\usepackage[usenames, dvipsnames]{color}
\usepackage{epstopdf}
\usepackage{bm}

\begin{document}
\title{\bf{Connectivity Percolation in Suspensions of Attractive Square-Well Spherocylinders}}

\author{Mohit Dixit}
\author{Hugues Meyer} 
\author{Tanja Schilling}
\affiliation{Research Unit for Physics and Materials Science, Universit\'{e} du Luxembourg, L-1511 Luxembourg, Luxembourg}

\maketitle

\begin{center}
  \large{\bf{Abstract}}\\
 \end{center}
We have studied the connectivity percolation transition in suspensions of attractive square-well spherocylinders by means of Monte Carlo simulation and connectednes percolation theory. In the 1980s the percolation threshold of slender fibres has been predicted to scale as the fibres' inverse aspect ratio (Phys.~Rev.~B {\bf 30}, 3933 (1984)). The main finding of our study is that the attractive spherocylinder system reaches this inverse scaling regime at much lower aspect ratios than found in suspensions of hard spherocylinders. We explain this difference by showing that third virial corrections of the pair connectedness functions, which are responsible for the deviation from the scaling regime, are less important for attractive potentials than for hard particles.
\clearpage

\section{Introduction} 

\addcontentsline{toc}{section}{Introduction} 

The connectivity percolation transition is the transition at which particles (or voids) connect up to form a system spanning network. The transport properties of disordered systems, such as e.g.~the electrical conductivity, depend sensitively on the concentration at which this transition occurs, i.e.~on the ``percolation threshold''.
The percolation threshold is determined by the shape of the particles and the interactions between them. Percolation of spherical particles with various repulsive and attractive interaction potentials has been studied in much detail \cite{Torquato:2002}. For non-spherical particles like fibres the effects of interactions have been studied less, despite the industrial interest in conducting fibres as fillers in conductive composites. 
Fibre shaped fillers have been analyzed theoretically \cite{Henderson:1967,Schoot:2008,Wang:2003, Chandler:1991}, in simulations with no interactions \cite{Pike:1974,Neda:1999,Foygel:2005,Mutiso:2012}, and in simulations with hard-core excluded volume interactions\cite{Sastry:2007,Tanja:2007,Tanja:2009,Grimaldi:2010}. In all of the systems the percolation threshold decreases if the aspect ratio of the fibres, i.e.~the length $L$ divided by the thickness $D$, increases. It has been conjectured that the percolation threshold should be proportional to $D/L$\cite{Balberg:1984}, but this relation is exact only for infinitely slender rods. In all of the systems listed above the scaling regime of slender rods is reached only for very large aspect ratios, larger than the aspect ratios of fibres that are commonly used as fillers in composite materials. Recently, van der Schoot et.~al.~rationalized this observation by means of connectedness percolation theory\cite{Tanja:2015} . 

As real fillers are often subject to attractive interactions, we present here a study on attractive square-well spherocylinders (SWSC) using Monte Carlo (MC) simulations as well as connectedness percolation theory. We discuss the scaling behaviour of the percolation threshold with aspect ratio and compare the results with the hard spherocylinders (HSC) system. 


\section{Methods}

We have performed MC simulations of hard spherocylinders with and without an attractive square-well potential. (We use the abbreviations SWSC for square-well attractive spherocylinder and HSC for hard spherocylinder for the rest of this text.) A spherocylinder consists of a cylinder of length $L$ and diameter $D$, capped with hemispheres of the same diameter. We have used a cubic simulation box with periodic boundary conditions. Configurations have been generated using single-particle displacements and rotations via the Metropolis scheme to sample the configurations space of the system\cite{Metropolis:1953}. We have generated configurations at fixed particle number $N$ and volume $V$, using simulation boxes of length $L_{x} = V^{1/3}$, where typically $L_{x} \approx 4L$  to $6L$.  
After equilibration, we generated 10,000 independent configurations for each value of $L/D$ and of the interaction parameters (which we will define below), to sample the probability that the system contains a percolating cluster. 
A special cell system \cite{Tanja:2005} has been employed for efficient overlap detection, where the box has been divided into a fine grid. This method is efficient for large aspect ratios, but expensive in terms of memory.
We have performed simulations of spherocylinders of aspect ratio $L/D$  ranging from 10 to 200. 

To define clusters of spherocylinders, a connectivity criterion is required. A pair of spherocylinders is said to be connected, if the line segments of the spherocylinders' axes are closer than a given value $\Delta D$, i.e.~a spherocylinder is surrounded by a contact shell of thickness $\lambda D = (\Delta - 1)D$ . When a cluster of connected spherocylinders wraps through the periodic boundaries the system percolates. We give the concentration of spherocylinders in terms of the volume fraction $\eta := Nv/V$, where $v =\pi D^{3}(2 + 3L/D)/12$ is the volume of the hard core of a spherocylinder. The volume fraction at the percolation threshold is called $\eta_{p}$. 

In an infinite system the percolation probability $p_{c}$ would rise instantaneously at the percolation threshold, but for a finite box size a sigmoidal curve is observed. The width of this curve decreases with increasing box size and its location shifts\cite{Stauffer:1985}. However, the volume fraction at which $p_{c}$ passes through 0.5 is almost independent of the box size. We therefore use this value to determine $\eta_{p}$. As we are interested in the qualitative behaviour and scaling properties of the percolation threshold, this rough criterion is sufficient. 

In the SWSC system, the spherocylinders interact via a square well potential with a width $\delta D$ and a depth $a\, k_{B}T$:

\begin{equation}
V(r)/k_{B}T = \left\{ 
  \begin{array}{l l}
    \infty & \quad \text{if $r$} < D\\
    -a & \quad \text{if } D \leq \text{$r$} \leq \delta D \\
    0 & \quad \text{if  $r$} > \delta D
  \end{array} \right.
\end{equation}

where $r$ is the axis to axis distance between two spherocylinders. For the HSC system $a = 0$.
  
This system can be interpreted as an extension of the Baxter hard sphere model\cite{Henderson:1967,baxter:1968} to rods. In the same spirit as for the Baxter spheres, we define a ``stickiness parameter'' $\tau$ 

\begin{equation}
\tau := \frac{1}{4(\delta ^{3}-1)(e^{a}-1)} \quad .
\end{equation}

The reduced second virial coefficient $B_{2}^{*}= B_{2}/B_{2}^{HS}$ is related to $\tau$ by

\begin{equation}
B_{2}^{*} = 1 - (\delta^{3} - 1)(e^{a} - 1) = 1 - 1/4\tau  \quad ,
\end{equation}

where $B_{2}^{HS} = 2\pi D^3/3$ is the second virial coefficient for hard spheres. The smaller the value of $\tau$ the more sticky are the particles\cite{Frenkel:2000,Miller:2004}. 

The attraction between two rods (eq.1) only depends on the 
surface-to-surface distance and not on their mutual orientation. Attractions 
between real fillers are usually either of the van der Waals 
type or caused by depletion. In both cases, the interaction strength depends 
on orientation (aligned rods attract each other more 
strongly than rods that lie perpendicular to each other). However, at 
the percolation threshold orientational correlations are weak\cite{Tanja:2007}, thus an interaction potential which does not depend on the angle should 
be sufficient to study percolation.

The exact functional form of the interaction potential will have an 
effect on the value of the percolation threshold, but the general trends 
that we discuss 
in the following for the square well potential should remain valid as 
long as the potential has no features that  significantly change 
the second and third order virial coefficients.


\section{Results and Discussion}

\begin{figure}[!htb]\centering
\includegraphics[width=0.45\linewidth]{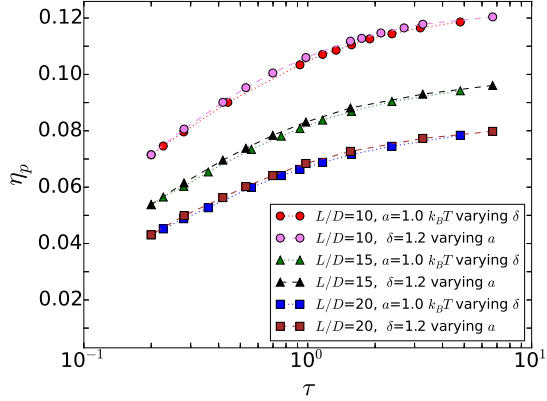}
\caption{ Percolation threshold $\eta_{p}$ versus stickiness $\tau$, varying $\delta$ and $a$ independently, for $L/D$ = 10,15 and 20. $\Delta = \delta$. $\eta_{p}$ decreases with both increasing stickiness (i.e.~decreasing $\tau$) and with increasing aspect ratio $L/D$. Curves for variation in $a$ and $\delta$ almost coincide, thus $\tau$ determines the percolation threshold. }
\label{fig:results5}
\end{figure}

We have investigated the percolation behaviour of suspensions of HSC and SWSC systems for varying aspect ratios  $L/D$  as well as interaction parameters  $a$ and $\delta$. To characterize the percolation transition, we have checked for invariant quantities at the percolation threshold. The number of contacts per spherocylinder turned out to be non-universal. The stickiness parameter $\tau$, however, is almost invariant, if one sets the connectivity range $\Delta D$ equal to the range of the potential $\delta D$.
In Figure \ref{fig:results5}, we have plotted the dependence of $\eta_{p}$ on the stickiness parameter $\tau$, varying $a$ and $\delta$ independently, for $L/D = 10, 15$ and $20$. The value of $a$ ranged between 0.05 and 1.0 and 
$\delta$ between 1.01 and 1.2. The percolation threshold values $\eta_{p}$ decrease as we increase the stickiness (i.e.~decrease the value of $\tau$). The curves for variation in $a$ and $\delta$ almost coincide for a particular aspect ratio $L/D$, which implies that $\tau$ (and thus $B_2$) is sufficient to have a good estimate of the percolation threshold for a particular aspect ratio.

\begin{figure}[!htb]\centering
\includegraphics[width=0.45\linewidth]{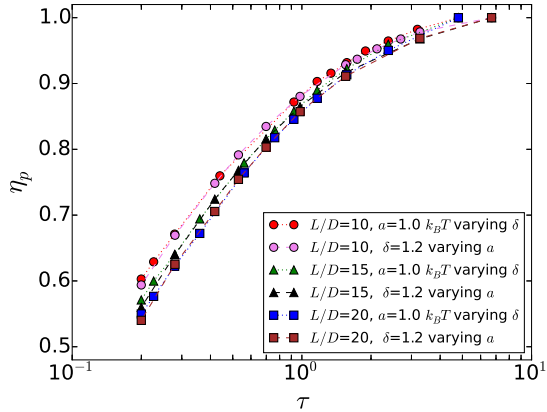}
\caption{ All the curves in Figure \ref{fig:results5} have been shifted to the highest value of $\eta_{p}$ for direct comparison. The comparative lowering in the percolation threshold $\eta_{p}$ with $\tau$ is higher as the aspect ratio $L/D$ is increased from 10 to 15, and 20.}
\label{fig:results6}
\end{figure}

To check for universal behaviour across aspect ratios $L/D$, we shift the curves in Figure \ref{fig:results5} to their highest $\eta_{p}$ values, see Figure \ref{fig:results6}. As evident from the graph, the dependence of $\eta_{p}$ on $\tau$ is not universal across aspect ratios. 

\begin{figure}[!htb]\centering
\includegraphics[width=0.45\linewidth]{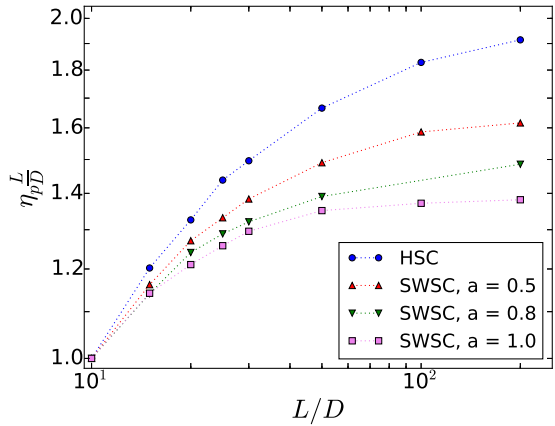}
\caption{To compare the scaling for large $L/D$ for HSC and SWSC systems,  $\eta_{p}\frac{L}{D}$ vs $L/D$ is plotted normalized by their respective $\eta_{p}*10$ values. $a = 0.5$ (triangle up), $a = 0.8$ (triangle down), and $a= 1.0$ (square), $\delta = \Delta = 1.2$  ($\tau = 0.53$, $\tau = 0.28$ and  $\tau = 0.2$, respectively). The SWSC system reaches inverse scaling at much smaller $L/D$ than the HSC system. HSC data from ref.~\cite{Tanja:2015}}
\label{fig:results8}
\end{figure}

Next we discuss the dependence of $\eta_{p}$ on the aspect ratio $L/D$ for the SWSC system and compare it to the HSC system. As we expect $\eta_{p} \propto D/L$ for large $L/D$  \cite{Balberg:1984}, we have plotted $\eta_{p}\frac{L}{D}$ in Figure \ref{fig:results8}, both for the SWSC systems as well as the HSC system. To allow for direct comparison, all curves have been shifted to the same value at $\frac{L}{D}=10$. The triangle up data points are for $a = 0.5$, the triangle down data points for $a=0.8$, the square data points for $a=1.0$, in all cases $\delta = 1.2$ ($\tau$ = 0.53, 0.28 and 0.2, respectively). All curves tend towards the slender rod limit in which $\eta_{p}$ scales as inverse with aspect ratio $L/D$. Surprisingly, however, the SWSC systems reaches the inverse scaling regime at much lower $L/D$ than the HSC system.

In suspensions of hard spherocylinders the inverse aspect ratio scaling regime is reached at very high aspect ratios ($L/D \gg 100$). This effects has recently been explained by van der Schoot et al. \cite{Tanja:2015} in the framework of connectedness percolation theory using the Parsons-Lee closure, which yields a density-dependent correction factor to the percolation threshold. Deviations from the inverse aspect ratio scaling for short rods come therefore from the expression of this correction factor, i.e.~indirectly from the Carnahan Starling equation of state, which enters the Parsons-Lee closure and includes a whole virial expansion. Since we do not know any accurate and convenient equation of state for square-well particles, we can not reproduce exactly this method in our specific case of attractive rods. However, we can go to a third order virial expansion in the general framework of percolation theory to explain the early inverse scaling of attractive spherocylinders.

According to classical connectivity percolation theory \cite{bug:1985,kyrylyuk:2008,coniglio:1977}, the overall mean cluster size $S$ is expressed as $S=1+\rho\hat{h}^{+}(q\rightarrow 0)$, where $\hat{(...)}$ stands for the 3-dimensional spatial Fourier Transform and the so-called total connectedness function $h^{+}(\bm{r},\bm{r}')$ is defined such that $\rho h^{+}(\bm{r},\bm{r}')d\bm{r}d\bm{r}'$ is the probability that two particles in volumes $d\bm{r}$ and $d\bm{r}'$ at positions $\bm{r}$ and $\bm{r}'$ are part of the same cluster. Inserting $h^{+}$ into a connectedness analogue of the Ornstein-Zernike (OZ) equation allows to define the direct pair connectedness function $C^{+}(\bm{r},\bm{r}')$. This definition through the connectedness OZ-equation yields $1+\rho\hat{h}^{+}=\left[ 1-\rho\hat{C}^{+} \right]^{-1}$, such that the mean cluster size $S$ diverges if $\rho \hat{C}^{+}(q\rightarrow 0)=1$. For practical purposes, and since we Fourier transform only at zero wave-vector, we will intentionally drop $(q\rightarrow 0)$ in the remaining text.

It has been shown \cite{coniglio:1977} that $C^{+}$ can be interpreted as the contribution of connected particles to the direct correlation function $C$, such that one can formally write $C=C^{+}+C^{*}$, where $C^{*}$ is the blocking part of the direct correlation function (DCF). Since the virial expansion of the DCF involves integrals of the Maier function $f(\bm{r})=\exp(-V(\bm{r})/k_{B}T)-1$, one also splits f into a connectedness part $f^{+}$ and a blocking part $f^{*}$ : $f=f^{+}+f^{*}$. In this framework, we have to distinguish two cases: either $\delta\leq\Delta$ or $\Delta\leq\delta$. In any case, $f^{+}(\bm{r}_{12})=\exp(-V(\bm{r})/k_{B}T)$ if $1$ and $2$ are connected but do not overlap and $f^{+}(\bm{r}_{12})=0$ otherwise. $f^{*}$ is then calculated as the difference between $f$ and $f^{+}$. By analogy with the virial expansion of the DCF, one can formally write $C^{+}$ as a virial expansion $C^{+} = \sum\limits_{n=2}^{\infty}\rho^{n-2}C_{n}^{+}$. Keeping only the first term yields $\rho_{c}=1/\hat{C}_{2}^{+}$, or $\phi_{c}=v/\hat{C}_{2}^{+}$ if expressed in terms of volume fraction. At the second virial level, $C^{+}_{2}=f^{+}$. Hence, $\hat{C}_{2}^{+} \propto \left[ e^{a}(\delta-1) + (\Delta-\delta) \right] D L^{2}$ if $\delta \leq \Delta$ or $\hat{C}_{2}^{+} \propto e^{a}(\Delta-1) D L^{2}$ if $\Delta \leq \delta$. Moreover $v\propto L D^{2}$ finally gives $\phi_{c}$ inversely proportional to the aspect ratio $l:=L/D$.

If we truncate the virial expansion at the third order, the percolation threshold is determined by $\rho_{c}\left( \hat{C}_{2}^{+} + \rho_{c} \hat{C}_{3}^{+} \right)=1$. One of the solutions of this equation is 
\begin{equation}
\rho_{c} = \frac{\hat{C}_{2}^{+}}{2\hat{C}_{3}^{+}}\left( \sqrt{1+4\frac{\hat{C}_{3}^{+}}{\hat{C}_{2}^{+2}}} -1 \right)
\end{equation}
If $|4\hat{C}_{3}^{+}/\hat{C}_{2}^{+2}| \ll 1$, we can Taylor expand the square root and we recover the second-virial solution $\rho_{c}=1/\hat{C}_{2}^{+}$.

\begin{figure}
\begin{center}
\includegraphics[width=.35\linewidth]{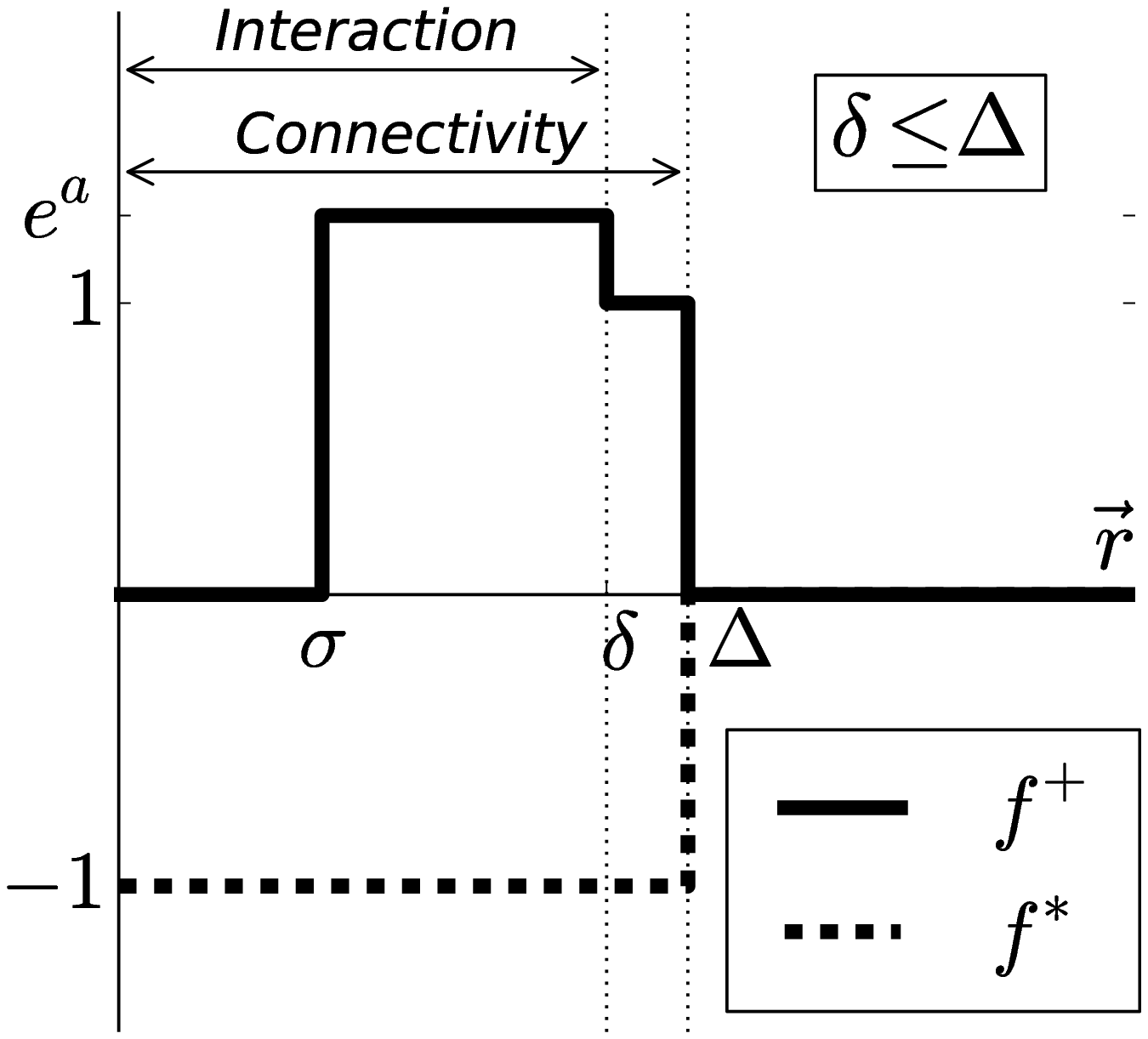}
\hspace*{.05\linewidth}
\includegraphics[width=.35\linewidth]{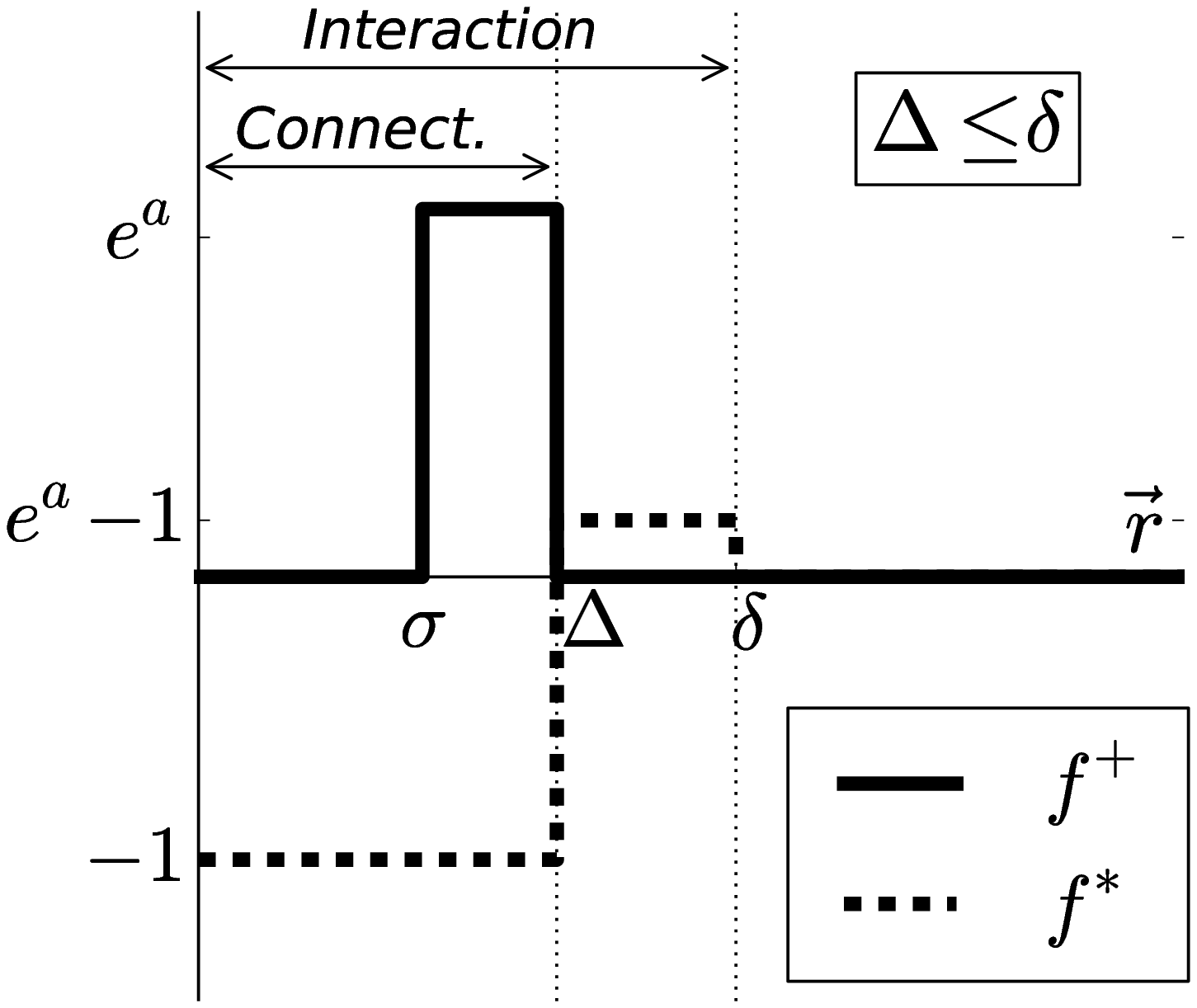}
\end{center}
\caption{$f^{+}$ and $f^{*}$ as a function of the interparticle distance $|\bm{r}|$. Connectedness and interaction ranges are indicated by the vertical dashed lines. Note that these functions are not identical whether $\Delta$ is greater or lower than $\delta$. However, in both cases, they can be expressed as sums of rectangular functions, which leads back to excluded volume considerations in the calculation of $C^{+}$.}
\label{fpfs}
\end{figure}

Coniglio showed that the third virial coefficient can be expanded as 
\begin{equation}
\begin{split}
\hat{C}_{3}^{+} = \int{\int{ d{\bf r}_{12} d{\bf r}_{13} \left( f^{+}_{12}f^{+}_{13}f^{+}_{23} +  f^{+}_{12}f^{+}_{13}f^{*}_{23} \right. }} \\ \left. + f^{+}_{12}f^{*}_{13}f^{+}_{23} + f^{*}_{12}f^{+}_{13}f^{+}_{23} + f^{+}_{12}f^{*}_{13}f^{*}_{23} \right) 
\end{split}
\end{equation}
corresponding to all diagrams for which $1$ and $2$ are directly or indirectly $f^{+}$-connected. One can formally rewrite it as $\hat{C}_{3}^{+} = I_{+++} + 3 I_{++*} + I_{+**}$ where $I$ refers the integrals with the corresponding number of $+$ and $*$. To compute these integrals, we recall that $f_{ij}^{+}$ and $f_{ij}^{*}$ are either 0 or constant, depending on the relative positions of $i$ and $j$ (see fig.\ref{fpfs}). Therefore, we write them as sums of rectangular functions of different widths and strengths and we compute the resulting integrals as three-body excluded volumes \cite{otten2011connectivity}, in which we keep the leading order terms which are proportional to $L^{3}$. We obtain 
\begin{equation}
\left\{
\begin{tabular}{l}
$I_{+++} \propto \left[ e^{a}(\delta-1) + (\Delta-\delta) \right]^{3} D^{3} L^{3}$ \vspace{5pt} \\
$I_{++*} \propto - \left[ e^{a}(\delta-1) + (\Delta-\delta) \right]^{2} \Delta D^{3} L^{3}$ \vspace{5pt} \\
$I_{+**} \propto \left[ e^{a}(\delta-1) + (\Delta-\delta) \right] \Delta^{2} D^{3} L^{3}$
\end{tabular}
\right., \text{    if } \delta \leq \Delta
\end{equation} 
\begin{equation}
\left\{
\begin{tabular}{l}
$I_{+++} \propto e^{3a}(\Delta-1)^{3} D^{3} L^{3}$ \vspace{5pt} \\
$I_{++*} \propto e^{2a}(\Delta-1)^{2}\left[ \left(e^{a}-1\right)(\delta-\Delta) - \Delta \right] D^{3} L^{3}$ \vspace{5pt} \\
$I_{+**} \propto e^{a}(\Delta-1)\left[ \left(e^{a}-1\right)(\delta-\Delta) - \Delta \right]^{2} D^{3} L^{3}$
\end{tabular}
\right., \text{    if } \Delta \leq \delta
\end{equation}
Note that these sets of expressions are consistent with each other for $\delta=\Delta$. We now focus on this particular case, in order to shrink the parameter space and to reduce the complexity of the calculations: $\delta=\Delta=:d$. We can therefore write
\begin{equation}
\hat{C}_{3}^{+} = \alpha e^{3a} (d-1)^{3} D^{3} L^{3} - 3\gamma e^{2a} (d-1)^{2} \Delta D^{3} L^{3} + \kappa e^{a} (d-1) \Delta^{2} D^{3} L^{3}
\end{equation}
where, $\alpha$, $\gamma$ and $\kappa$ are constants. 
Since the integrals $I$ are computed using excluded volume considerations, all terms are of purely geometric origin. More precisely, they consist of combinations of geometric intersections of objects of the same shape but with different dimensions ($\Delta-\delta$, $\delta-1$, ...). Therefore the prefactors in all these calculations have to be the same (and of the order of $\pi^{2}$), and we can reasonably assume that $\alpha\approx\gamma\approx\kappa\approx 1$. Moreover, $C_{2}^{+}$ consists also of 2-body excluded volumes; therefore the prefactor involved in the leading order of $C_{2}^{+2}$ should also be close to the one of the leading order of a 3-body excluded volume, namely $\alpha$. Thus, we neglect all these prefactors in the ratio $K:=\left| 4\hat{C}_{3}^{+}/\hat{C}_{2}^{+2} \right|$. It needs to be as small as possible in order to reach the $D/L$ scaling. This approximation may be a bit rough but would only lead to a change of a global prefactor, which does not influence the rest of our argument. We can write now
\begin{equation}
K(a) = \frac{4d}{l} \left| \mu(a) - 3 + \frac{1}{\mu(a)}  \right|
\end{equation}
where $\mu(a)=\left(1-\frac{1}{d}\right)e^{a}$.
First of all, we notice that for infinite aspect ratio $l$, $K$ vanishes such that we are in the second virial limit. In addition, if $\mu(a) \ll 1$, the first term is negligible and we have a decreasing function of $a$, at least for small values of attraction strength. Since working with too large values of $a$ would require a virial expansion to the fourth order, $K$ is nevertheless always lower than $1$ in the range of validity.

\begin{figure}
\begin{center}
\includegraphics[width=.45\linewidth]{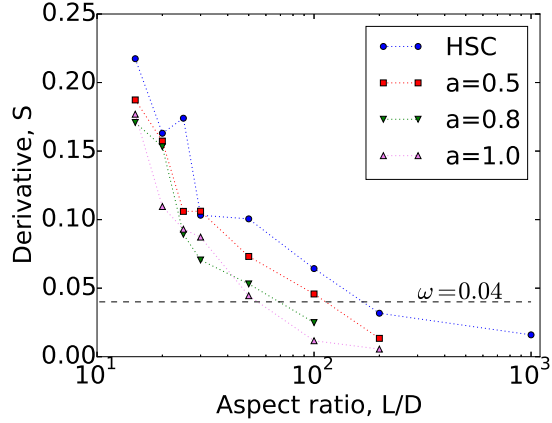}
\end{center}
\caption{ Derivative of the curve from Fig.~\ref{fig:results8}. The values have been evaluated by connecting each consecutive pair of data points by a straight line. The criterion $\omega$ has been set to 0.04 shown in the graph by a horizontal dashed line.}
\label{fig:results1}
\end{figure}
\begin{figure}
\begin{center}
\includegraphics[width=.45\linewidth]{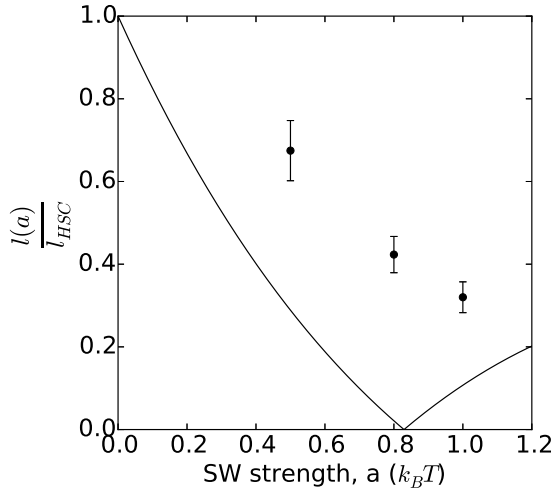}
\end{center}
\caption{ The relative importance of the third virial coefficient is always smaller for SWSC systems as compared to the HSC, which makes the square root correction less important for the percolation threshold.}
\label{fig:results9}
\end{figure}

Let us define a threshold value $\omega$ aimed at setting a criterion which determines if the system is in the second-virial scaling. For a particular interaction strength $a$, we compute the aspect ratio $l(a)$ such that $K(a,l(a))=\omega$. Any apsect ratio $l'>l(a)$ will be such that  $K(a,l')<\omega$. If $\omega$ is chosen sufficiently small, this indicates that $K(a,l')$ is also small enough to Taylor expand equation (5) so that the second-virial scaling is reached : $l(a)$ is therefore the minimal aspect ratio for which this asymptotic behaviour is obtained. Thus, using equation (9), and chosing the same threshold value $\omega$ for any strength $a$, one has
\begin{equation}
\frac{l(a)}{l_{HSC}}=  \left| \frac{\mu(a) - 3 + \frac{1}{\mu(a)}}{ \frac{d}{d-1} - \frac{1}{d} - 2} \right|
\end{equation}
where $l_{HSC}$ indicates the minimal aspect ratio for which the scaling limit is reached in the case of hard sphecoylinders, i.e. $a=0$. This function is plotted in Fig. \ref{fig:results9} for $d=1.2$. We notice a non-monotonic behaviour as well as a particular interaction strength for which $l=0$, suggesting that the scaling regime is obtained from the sphere on, for this particular strength. This is not obvious and is not supported by the simulation data. Considering higher virial orders should very probably cancel this effect. Moreover, prefactors have been neglected in our study. An exact calculation would certainly improve the theoretical curve although such a calculation would require a huge amount of work.

In order to test the validity of this analysis, we approximate the derivative $S$ of the simulation data in Fig.~\ref{fig:results8} by the slope of a straight line through each consecutive pair of data points
\begin{equation}
S_{i+1} = \frac{{(\eta_{p}\frac{L}{D})}_{i+1} -{(\eta_{p}\frac{L}{D})}_{i}}{(\frac{L}{D})_{i+1} - (\frac{L}{D})_{i}} \quad .
\end{equation} This derivative vanishes in the inverse scaling regime. For each square well depth $a$, we identify the aspect ratio for which this derivative becomes smaller than an arbitrary, small value taken as 0.04 shown in Fig.~\ref{fig:results1}. This criterion can be compared to the criterion $\omega$ mentioned in the previous paragraph. The simulation points have been superimposed on the theoretical prediction in Fig.~\ref{fig:results9}. Although the ratio $\frac{l(a)}{l_{HSC}}$ does not depend on $\omega$ in the theoretical analysis, it actually strongly depends on the small parameter used to evaluate the simulation data, which leads to very large errorbars. Since the theory is based on strong approximations, the agreement of the $\frac{l(a)}{l_{HSC}}$ value with the simulation is not as important as the trend that is observed. Our theoretical argument together with the simulation data shows that "stickiness" between spherocylinders reduces deviations from the inverse scaling regime.

\section{Conclusion}

We have investigated percolation in suspensions of attractive square well spherocylinders by means of computer simulations and connectedness percolation theory. The main finding is that SWSCs reach the regime in which the percolation threshold scales as the inverse aspect ratio at much shorter aspect ratios than hard spherocylinders. The more sticky the spherocylinders, the smaller the aspect ratio at which scaling is reached. On a third virial level there even seems to be a value of stickiness for which $D/L$ scaling starts already at the limit of spheres. We also find that the stickiness parameter at percolation is almost invariant across the parameter space of the potential for a particular aspect ratio.  

\section{Acknowledgments} 

We thank Paul van der Schoot for stimulating discussions. This project was financially supported by the Fonds National de la recherche Luxembourg within the DFG-FNR INTER project ``Thin Film Growth''. Data from computer simulations presented in this paper were produced using the HPC facilities of University of Luxembourg\cite{HPC}.

\bibliographystyle{apsrev4-1} 

\begin{thebibliography}{27}%
\makeatletter
\providecommand \@ifxundefined [1]{%
 \@ifx{#1\undefined}
}%
\providecommand \@ifnum [1]{%
 \ifnum #1\expandafter \@firstoftwo
 \else \expandafter \@secondoftwo
 \fi
}%
\providecommand \@ifx [1]{%
 \ifx #1\expandafter \@firstoftwo
 \else \expandafter \@secondoftwo
 \fi
}%
\providecommand \natexlab [1]{#1}%
\providecommand \enquote  [1]{``#1''}%
\providecommand \bibnamefont  [1]{#1}%
\providecommand \bibfnamefont [1]{#1}%
\providecommand \citenamefont [1]{#1}%
\providecommand \href@noop [0]{\@secondoftwo}%
\providecommand \href [0]{\begingroup \@sanitize@url \@href}%
\providecommand \@href[1]{\@@startlink{#1}\@@href}%
\providecommand \@@href[1]{\endgroup#1\@@endlink}%
\providecommand \@sanitize@url [0]{\catcode `\\12\catcode `\$12\catcode
  `\&12\catcode `\#12\catcode `\^12\catcode `\_12\catcode `\%12\relax}%
\providecommand \@@startlink[1]{}%
\providecommand \@@endlink[0]{}%
\providecommand \url  [0]{\begingroup\@sanitize@url \@url }%
\providecommand \@url [1]{\endgroup\@href {#1}{\urlprefix }}%
\providecommand \urlprefix  [0]{URL }%
\providecommand \Eprint [0]{\href }%
\providecommand \doibase [0]{http://dx.doi.org/}%
\providecommand \selectlanguage [0]{\@gobble}%
\providecommand \bibinfo  [0]{\@secondoftwo}%
\providecommand \bibfield  [0]{\@secondoftwo}%
\providecommand \translation [1]{[#1]}%
\providecommand \BibitemOpen [0]{}%
\providecommand \bibitemStop [0]{}%
\providecommand \bibitemNoStop [0]{.\EOS\space}%
\providecommand \EOS [0]{\spacefactor3000\relax}%
\providecommand \BibitemShut  [1]{\csname bibitem#1\endcsname}%
\let\auto@bib@innerbib\@empty
\bibitem [{\citenamefont {Torquato}(2002)}]{Torquato:2002}%
  \BibitemOpen
  \bibfield  {author} {\bibinfo {author} {\bibfnamefont {S.}~\bibnamefont
  {Torquato}},\ }\href@noop {} {\bibfield  {journal} {\bibinfo  {journal}
  {Springer}\ }\textbf {\bibinfo {volume} {16}} (\bibinfo {year}
  {2002})}\BibitemShut {NoStop}%
\bibitem [{\citenamefont {Barker}\ and\ \citenamefont
  {Henderson}(1967)}]{Henderson:1967}%
  \BibitemOpen
  \bibfield  {author} {\bibinfo {author} {\bibfnamefont {J.~A.}\ \bibnamefont
  {Barker}}\ and\ \bibinfo {author} {\bibfnamefont {D.}~\bibnamefont
  {Henderson}},\ }\href@noop {} {\bibfield  {journal} {\bibinfo  {journal} {The
  Journal of chemical physics}\ }\textbf {\bibinfo {volume} {47}},\ \bibinfo
  {pages} {4714} (\bibinfo {year} {1967})}\BibitemShut {NoStop}%
\bibitem [{\citenamefont {AV}\ and\ \citenamefont {van~der
  Schoot~P}(2008)}]{Schoot:2008}%
  \BibitemOpen
  \bibfield  {author} {\bibinfo {author} {\bibfnamefont {K.}~\bibnamefont
  {AV}}\ and\ \bibinfo {author} {\bibnamefont {van~der Schoot~P}},\ }\href@noop
  {} {\bibfield  {journal} {\bibinfo  {journal} {Proc. Natl. Acad. Sci. USA}\
  }\textbf {\bibinfo {volume} {105}},\ \bibinfo {pages} {8221} (\bibinfo {year}
  {2008})}\BibitemShut {NoStop}%
\bibitem [{\citenamefont {Wang}\ and\ \citenamefont
  {Chatterjee}(2003)}]{Wang:2003}%
  \BibitemOpen
  \bibfield  {author} {\bibinfo {author} {\bibfnamefont {X.}~\bibnamefont
  {Wang}}\ and\ \bibinfo {author} {\bibfnamefont {A.~P.}\ \bibnamefont
  {Chatterjee}},\ }\href@noop {} {\bibfield  {journal} {\bibinfo  {journal}
  {The Journal of Chemical Physics}\ }\textbf {\bibinfo {volume} {118}},\
  \bibinfo {pages} {10787} (\bibinfo {year} {2003})}\BibitemShut {NoStop}%
\bibitem [{\citenamefont {Leung}\ and\ \citenamefont
  {Chandler}(1991)}]{Chandler:1991}%
  \BibitemOpen
  \bibfield  {author} {\bibinfo {author} {\bibfnamefont {K.}~\bibnamefont
  {Leung}}\ and\ \bibinfo {author} {\bibfnamefont {D.}~\bibnamefont
  {Chandler}},\ }\href@noop {} {\bibfield  {journal} {\bibinfo  {journal}
  {Journal of Statistical Physics}\ }\textbf {\bibinfo {volume} {63}},\
  \bibinfo {pages} {837} (\bibinfo {year} {1991})}\BibitemShut {NoStop}%
\bibitem [{\citenamefont {Pike}\ and\ \citenamefont
  {Seager}(1974)}]{Pike:1974}%
  \BibitemOpen
  \bibfield  {author} {\bibinfo {author} {\bibfnamefont {G.~E.}\ \bibnamefont
  {Pike}}\ and\ \bibinfo {author} {\bibfnamefont {C.~H.}\ \bibnamefont
  {Seager}},\ }\href@noop {} {\bibfield  {journal} {\bibinfo  {journal} {Phys.
  Rev. B}\ }\textbf {\bibinfo {volume} {10}},\ \bibinfo {pages} {1421}
  (\bibinfo {year} {1974})}\BibitemShut {NoStop}%
\bibitem [{\citenamefont {N\'eda}\ \emph {et~al.}(1999)\citenamefont {N\'eda},
  \citenamefont {Florian},\ and\ \citenamefont {Brechet}}]{Neda:1999}%
  \BibitemOpen
  \bibfield  {author} {\bibinfo {author} {\bibfnamefont {Z.}~\bibnamefont
  {N\'eda}}, \bibinfo {author} {\bibfnamefont {R.}~\bibnamefont {Florian}}, \
  and\ \bibinfo {author} {\bibfnamefont {Y.}~\bibnamefont {Brechet}},\
  }\href@noop {} {\bibfield  {journal} {\bibinfo  {journal} {Phys. Rev. E}\
  }\textbf {\bibinfo {volume} {59}},\ \bibinfo {pages} {3717} (\bibinfo {year}
  {1999})}\BibitemShut {NoStop}%
\bibitem [{\citenamefont {Foygel}\ \emph {et~al.}(2005)\citenamefont {Foygel},
  \citenamefont {Morris}, \citenamefont {Anez}, \citenamefont {French},\ and\
  \citenamefont {Sobolev}}]{Foygel:2005}%
  \BibitemOpen
  \bibfield  {author} {\bibinfo {author} {\bibfnamefont {M.}~\bibnamefont
  {Foygel}}, \bibinfo {author} {\bibfnamefont {R.~D.}\ \bibnamefont {Morris}},
  \bibinfo {author} {\bibfnamefont {D.}~\bibnamefont {Anez}}, \bibinfo {author}
  {\bibfnamefont {S.}~\bibnamefont {French}}, \ and\ \bibinfo {author}
  {\bibfnamefont {V.~L.}\ \bibnamefont {Sobolev}},\ }\href@noop {} {\bibfield
  {journal} {\bibinfo  {journal} {Phys. Rev. B}\ }\textbf {\bibinfo {volume}
  {71}},\ \bibinfo {pages} {104201} (\bibinfo {year} {2005})}\BibitemShut
  {NoStop}%
\bibitem [{\citenamefont {Mutiso}\ \emph {et~al.}(2012)\citenamefont {Mutiso},
  \citenamefont {Sherrott}, \citenamefont {Li},\ and\ \citenamefont
  {Winey}}]{Mutiso:2012}%
  \BibitemOpen
  \bibfield  {author} {\bibinfo {author} {\bibfnamefont {R.~M.}\ \bibnamefont
  {Mutiso}}, \bibinfo {author} {\bibfnamefont {M.~C.}\ \bibnamefont
  {Sherrott}}, \bibinfo {author} {\bibfnamefont {J.}~\bibnamefont {Li}}, \ and\
  \bibinfo {author} {\bibfnamefont {K.~I.}\ \bibnamefont {Winey}},\ }\href@noop
  {} {\bibfield  {journal} {\bibinfo  {journal} {Phys. Rev. B}\ }\textbf
  {\bibinfo {volume} {86}},\ \bibinfo {pages} {214306} (\bibinfo {year}
  {2012})}\BibitemShut {NoStop}%
\bibitem [{\citenamefont {Berhan}\ and\ \citenamefont
  {Sastry}(2007)}]{Sastry:2007}%
  \BibitemOpen
  \bibfield  {author} {\bibinfo {author} {\bibfnamefont {L.}~\bibnamefont
  {Berhan}}\ and\ \bibinfo {author} {\bibfnamefont {A.~M.}\ \bibnamefont
  {Sastry}},\ }\href@noop {} {\bibfield  {journal} {\bibinfo  {journal} {Phys.
  Rev. E}\ }\textbf {\bibinfo {volume} {75}},\ \bibinfo {pages} {041121}
  (\bibinfo {year} {2007})}\BibitemShut {NoStop}%
\bibitem [{\citenamefont {Schilling}\ \emph {et~al.}(2007)\citenamefont
  {Schilling}, \citenamefont {Jungblut},\ and\ \citenamefont
  {Miller}}]{Tanja:2007}%
  \BibitemOpen
  \bibfield  {author} {\bibinfo {author} {\bibfnamefont {T.}~\bibnamefont
  {Schilling}}, \bibinfo {author} {\bibfnamefont {S.}~\bibnamefont {Jungblut}},
  \ and\ \bibinfo {author} {\bibfnamefont {M.~A.}\ \bibnamefont {Miller}},\
  }\href@noop {} {\bibfield  {journal} {\bibinfo  {journal} {Phys. Rev. Lett.}\
  }\textbf {\bibinfo {volume} {98}},\ \bibinfo {pages} {108303} (\bibinfo
  {year} {2007})}\BibitemShut {NoStop}%
\bibitem [{\citenamefont {Schilling}\ \emph {et~al.}(2010)\citenamefont
  {Schilling}, \citenamefont {Jungblut},\ and\ \citenamefont
  {Miller}}]{Tanja:2009}%
  \BibitemOpen
  \bibfield  {author} {\bibinfo {author} {\bibfnamefont {T.}~\bibnamefont
  {Schilling}}, \bibinfo {author} {\bibfnamefont {S.}~\bibnamefont {Jungblut}},
  \ and\ \bibinfo {author} {\bibfnamefont {M.~A.}\ \bibnamefont {Miller}},\
  }\href@noop {} {\bibfield  {journal} {\bibinfo  {journal} {Taylor and
  Francis}\ } (\bibinfo {year} {2010})}\BibitemShut {NoStop}%
\bibitem [{\citenamefont {Ambrosetti}\ \emph {et~al.}(2010)\citenamefont
  {Ambrosetti}, \citenamefont {Grimaldi}, \citenamefont {Balberg},
  \citenamefont {Maeder}, \citenamefont {Danani},\ and\ \citenamefont
  {Ryser}}]{Grimaldi:2010}%
  \BibitemOpen
  \bibfield  {author} {\bibinfo {author} {\bibfnamefont {G.}~\bibnamefont
  {Ambrosetti}}, \bibinfo {author} {\bibfnamefont {C.}~\bibnamefont
  {Grimaldi}}, \bibinfo {author} {\bibfnamefont {I.}~\bibnamefont {Balberg}},
  \bibinfo {author} {\bibfnamefont {T.}~\bibnamefont {Maeder}}, \bibinfo
  {author} {\bibfnamefont {A.}~\bibnamefont {Danani}}, \ and\ \bibinfo {author}
  {\bibfnamefont {P.}~\bibnamefont {Ryser}},\ }\href@noop {} {\bibfield
  {journal} {\bibinfo  {journal} {Phys. Rev. B}\ }\textbf {\bibinfo {volume}
  {81}},\ \bibinfo {pages} {155434} (\bibinfo {year} {2010})}\BibitemShut
  {NoStop}%
\bibitem [{\citenamefont {Balberg}\ \emph {et~al.}(1984)\citenamefont
  {Balberg}, \citenamefont {Anderson}, \citenamefont {Alexander},\ and\
  \citenamefont {Wagner}}]{Balberg:1984}%
  \BibitemOpen
  \bibfield  {author} {\bibinfo {author} {\bibfnamefont {I.}~\bibnamefont
  {Balberg}}, \bibinfo {author} {\bibfnamefont {C.}~\bibnamefont {Anderson}},
  \bibinfo {author} {\bibfnamefont {S.}~\bibnamefont {Alexander}}, \ and\
  \bibinfo {author} {\bibfnamefont {N.}~\bibnamefont {Wagner}},\ }\href@noop {}
  {\bibfield  {journal} {\bibinfo  {journal} {Physical review B}\ }\textbf
  {\bibinfo {volume} {30}},\ \bibinfo {pages} {3933} (\bibinfo {year}
  {1984})}\BibitemShut {NoStop}%
\bibitem [{\citenamefont {Schilling}\ \emph {et~al.}(2015)\citenamefont
  {Schilling}, \citenamefont {Miller},\ and\ \citenamefont {van~der
  Schoot}}]{Tanja:2015}%
  \BibitemOpen
  \bibfield  {author} {\bibinfo {author} {\bibfnamefont {T.}~\bibnamefont
  {Schilling}}, \bibinfo {author} {\bibfnamefont {M.}~\bibnamefont {Miller}}, \
  and\ \bibinfo {author} {\bibfnamefont {P.}~\bibnamefont {van~der Schoot}},\
  }\href@noop {} {\bibfield  {journal} {\bibinfo  {journal} {Europhys. Lett.}\
  }\textbf {\bibinfo {volume} {111}},\ \bibinfo {pages} {56004} (\bibinfo
  {year} {May 2015})}\BibitemShut {NoStop}%
\bibitem [{\citenamefont {Metropolis}\ \emph {et~al.}(1953)\citenamefont
  {Metropolis}, \citenamefont {Rosenbluth}, \citenamefont {Rosenbluth},
  \citenamefont {Teller},\ and\ \citenamefont {Teller}}]{Metropolis:1953}%
  \BibitemOpen
  \bibfield  {author} {\bibinfo {author} {\bibfnamefont {N.}~\bibnamefont
  {Metropolis}}, \bibinfo {author} {\bibfnamefont {A.~W.}\ \bibnamefont
  {Rosenbluth}}, \bibinfo {author} {\bibfnamefont {M.~N.}\ \bibnamefont
  {Rosenbluth}}, \bibinfo {author} {\bibfnamefont {A.~H.}\ \bibnamefont
  {Teller}}, \ and\ \bibinfo {author} {\bibfnamefont {E.}~\bibnamefont
  {Teller}},\ }\href@noop {} {\bibfield  {journal} {\bibinfo  {journal} {The
  Journal of Chemical Physics}\ }\textbf {\bibinfo {volume} {21}},\ \bibinfo
  {pages} {1087} (\bibinfo {year} {1953})}\BibitemShut {NoStop}%
\bibitem [{\citenamefont {Vink}\ and\ \citenamefont
  {Schilling}(2005)}]{Tanja:2005}%
  \BibitemOpen
  \bibfield  {author} {\bibinfo {author} {\bibfnamefont {R.~L.~C.}\
  \bibnamefont {Vink}}\ and\ \bibinfo {author} {\bibfnamefont {T.}~\bibnamefont
  {Schilling}},\ }\href@noop {} {\bibfield  {journal} {\bibinfo  {journal}
  {Phys. Rev. E}\ }\textbf {\bibinfo {volume} {71}},\ \bibinfo {pages} {051716}
  (\bibinfo {year} {2005})}\BibitemShut {NoStop}%
\bibitem [{\citenamefont {Aharony}\ and\ \citenamefont
  {Stauffer}(2003)}]{Stauffer:1985}%
  \BibitemOpen
  \bibfield  {author} {\bibinfo {author} {\bibfnamefont {A.}~\bibnamefont
  {Aharony}}\ and\ \bibinfo {author} {\bibfnamefont {D.}~\bibnamefont
  {Stauffer}},\ }\href@noop {} {\bibfield  {journal} {\bibinfo  {journal}
  {Taylor $\&$ Francis}\ } (\bibinfo {year} {2003})}\BibitemShut {NoStop}%
\bibitem [{\citenamefont {Nigro}\ and\ \citenamefont
  {Grimaldi}(2014)}]{Nigro:2014}%
  \BibitemOpen
  \bibfield  {author} {\bibinfo {author} {\bibfnamefont {B.}~\bibnamefont
  {Nigro}}\ and\ \bibinfo {author} {\bibfnamefont {C.}~\bibnamefont
  {Grimaldi}},\ }\href@noop {} {\bibfield  {journal} {\bibinfo  {journal}
  {Phys. Rev. B}\ }\textbf {\bibinfo {volume} {90}},\ \bibinfo {pages} {094202}
  (\bibinfo {year} {2014})}\BibitemShut {NoStop}%
\bibitem [{\citenamefont {Baxter}(1968)}]{baxter:1968}%
  \BibitemOpen
  \bibfield  {author} {\bibinfo {author} {\bibfnamefont {R.}~\bibnamefont
  {Baxter}},\ }\href@noop {} {\bibfield  {journal} {\bibinfo  {journal} {The
  Journal of Chemical Physics}\ }\textbf {\bibinfo {volume} {49}},\ \bibinfo
  {pages} {2770} (\bibinfo {year} {1968})}\BibitemShut {NoStop}%
\bibitem [{\citenamefont {Noro}\ and\ \citenamefont
  {Frenkel}(2000)}]{Frenkel:2000}%
  \BibitemOpen
  \bibfield  {author} {\bibinfo {author} {\bibfnamefont {M.~G.}\ \bibnamefont
  {Noro}}\ and\ \bibinfo {author} {\bibfnamefont {D.}~\bibnamefont {Frenkel}},\
  }\href@noop {} {\bibfield  {journal} {\bibinfo  {journal} {The Journal of
  Chemical Physics}\ }\textbf {\bibinfo {volume} {113}},\ \bibinfo {pages}
  {2941} (\bibinfo {year} {2000})}\BibitemShut {NoStop}%
\bibitem [{\citenamefont {Miller}\ and\ \citenamefont
  {Frenkel}(2004)}]{Miller:2004}%
  \BibitemOpen
  \bibfield  {author} {\bibinfo {author} {\bibfnamefont {M.~A.}\ \bibnamefont
  {Miller}}\ and\ \bibinfo {author} {\bibfnamefont {D.}~\bibnamefont
  {Frenkel}},\ }\href@noop {} {\bibfield  {journal} {\bibinfo  {journal} {The
  Journal of chemical physics}\ }\textbf {\bibinfo {volume} {121}},\ \bibinfo
  {pages} {535} (\bibinfo {year} {2004})}\BibitemShut {NoStop}%
\bibitem [{\citenamefont {Bug}\ \emph {et~al.}(1985)\citenamefont {Bug},
  \citenamefont {Safran},\ and\ \citenamefont {Webman}}]{bug:1985}%
  \BibitemOpen
  \bibfield  {author} {\bibinfo {author} {\bibfnamefont {A.}~\bibnamefont
  {Bug}}, \bibinfo {author} {\bibfnamefont {S.}~\bibnamefont {Safran}}, \ and\
  \bibinfo {author} {\bibfnamefont {I.}~\bibnamefont {Webman}},\ }\href@noop {}
  {\bibfield  {journal} {\bibinfo  {journal} {Physical review letters}\
  }\textbf {\bibinfo {volume} {54}},\ \bibinfo {pages} {1412} (\bibinfo {year}
  {1985})}\BibitemShut {NoStop}%
\bibitem [{\citenamefont {Kyrylyuk}\ and\ \citenamefont {van~der
  Schoot}(2008)}]{kyrylyuk:2008}%
  \BibitemOpen
  \bibfield  {author} {\bibinfo {author} {\bibfnamefont {A.~V.}\ \bibnamefont
  {Kyrylyuk}}\ and\ \bibinfo {author} {\bibfnamefont {P.}~\bibnamefont {van~der
  Schoot}},\ }\href@noop {} {\bibfield  {journal} {\bibinfo  {journal}
  {Proceedings of the National Academy of Sciences}\ }\textbf {\bibinfo
  {volume} {105}},\ \bibinfo {pages} {8221} (\bibinfo {year}
  {2008})}\BibitemShut {NoStop}%
\bibitem [{\citenamefont {Coniglio}\ \emph {et~al.}(1977)\citenamefont
  {Coniglio}, \citenamefont {Angelis},\ and\ \citenamefont
  {Forlani}}]{coniglio:1977}%
  \BibitemOpen
  \bibfield  {author} {\bibinfo {author} {\bibfnamefont {A.}~\bibnamefont
  {Coniglio}}, \bibinfo {author} {\bibfnamefont {U.~D.}\ \bibnamefont
  {Angelis}}, \ and\ \bibinfo {author} {\bibfnamefont {A.}~\bibnamefont
  {Forlani}},\ }\href@noop {} {\bibfield  {journal} {\bibinfo  {journal}
  {Journal of Physics A: Mathematical and General}\ }\textbf {\bibinfo {volume}
  {10}},\ \bibinfo {pages} {1123} (\bibinfo {year} {1977})}\BibitemShut
  {NoStop}%
\bibitem [{\citenamefont {Otten}\ and\ \citenamefont {van~der
  Schoot}(2011)}]{otten2011connectivity}%
  \BibitemOpen
  \bibfield  {author} {\bibinfo {author} {\bibfnamefont {R.~H.}\ \bibnamefont
  {Otten}}\ and\ \bibinfo {author} {\bibfnamefont {P.}~\bibnamefont {van~der
  Schoot}},\ }\href@noop {} {\bibfield  {journal} {\bibinfo  {journal} {The
  Journal of chemical physics}\ }\textbf {\bibinfo {volume} {134}},\ \bibinfo
  {pages} {094902} (\bibinfo {year} {2011})}\BibitemShut {NoStop}%
\bibitem [{\citenamefont {Varrette}\ \emph {et~al.}(2014)\citenamefont
  {Varrette}, \citenamefont {Bouvry}, \citenamefont {Cartiaux},\ and\
  \citenamefont {Georgatos}}]{HPC}%
  \BibitemOpen
  \bibfield  {author} {\bibinfo {author} {\bibfnamefont {S.}~\bibnamefont
  {Varrette}}, \bibinfo {author} {\bibfnamefont {P.}~\bibnamefont {Bouvry}},
  \bibinfo {author} {\bibfnamefont {H.}~\bibnamefont {Cartiaux}}, \ and\
  \bibinfo {author} {\bibfnamefont {F.}~\bibnamefont {Georgatos}},\ }\href@noop
  {} {\bibfield  {journal} {\bibinfo  {journal} {IEEE, Bologna, Italy}\ }
  (\bibinfo {year} {2014})}\BibitemShut {NoStop}%
\end{thebibliography}
\end{document}